\documentclass[aps,preprint]{revtex4}
\usepackage{amsmath}
\usepackage{epsfig}

\begin{document}
\title{Case study: calculation of a narrow resonance\\ with the LIT method }

\author{Winfried Leidemann$^{1,2}$} 
\affiliation{$^{1}$Dipartimento di Fisica, Universit\`a di Trento,
  I-38100 Trento, Italy\\
$^{2}$Istituto Nazionale di Fisica Nucleare, Gruppo Collegato
  di Trento, Italy\\
}

\date{\today}

\begin{abstract}
The possibility to resolve narrow structures in reaction cross sections in calculations with 
the Lorentz integral transform (LIT) method is studied. To this end we consider a fictitious
two-nucleon problem with a low-lying and narrow resonance in the $^3P_1$ nucleon-nucleon
partial wave and calculate the corresponding ``deuteron photoabsorption cross section''.
In the LIT method the use of continuum wave functions is avoided and one works
instead with a localized function $\tilde\Psi$. In this case study it is investigated how 
far into the asymptotic region $\tilde\Psi$ has to be determined in order to obtain a 
precise resolution of the artificially introduced E1 resonance. Comparing with the
results of a conventional calculation with explicit neutron-proton continuum wave functions
it is shown that the LIT approach leads to an excellent reproduction of the cross section 
in the resonance region and of further finer cross section details at higher energies.
To this end, however, for $\tilde\Psi$ one has to take into account two-nucleon distances 
up to at least 30 fm.
\end{abstract}

\bigskip

\maketitle
\section{Introduction}

The LIT approach \cite{ELO94} allows the ab initio calculation of reaction cross sections,
where a many-body continuum is involved. The great advantage of the LIT method
lies in the fact that the knowledge of the generally complicated many-body 
continuum wave function is not required. In fact the scattering problem 
is reduced to a calculation of a localized function with an asymptotic boundary condition
similar to a bound-state wave function. The LIT method has been applied to various
electroweak cross sections in the nuclear mass range from A=3 to A=7. Among the applications 
are the first realistic ab initio calculations of the nuclear three- and four-body total
photoabsorption cross sections \cite{ELOT00,Doron06} as well as of the inelastic neutral current 
neutrino scattering off $^4$He \cite{GN07}. In addition first ab initio calculations were 
performed for the total photoabsorption cross sections of $^{4,6}$He and $^{6,7}$Li with 
semirealistic forces \cite{4-body,6-body,7-body}. Other applications were carried out for the 
inelastic inclusive electron scattering cross section (see e.g. \cite{ee'97,ee'05,ee'08}).
Further applications and a detailed description of the LIT method are presented
in a recent review article \cite{ELOB07}.

In the past the LIT technique was mainly applied to cases where one has typically rather broad 
structures (quasielastic peak, giant dipole resonance). One may ask what is the power of the 
method in presence of narrow resonances in the continuum. In a recent LIT calculation of the 
(e,e') longitudinal and transverse form factors of $^4$He \cite{Sonia07} a resonance in the 
Coulomb monopole transition was observed but the width of the resonance could not be determined. 
Therefore in the present paper we want to address this problem and investigate how narrow
resonances can be calculated with the LIT method. To this end we consider a fictitious
two-nucleon problem with a low-lying and narrow resonance in the $^3P_1$ nucleon-nucleon
partial wave and calculate the corresponding ``deuteron photoabsorption cross section'', which
exhibits a pronounced E1 resonance at a photon energy of 2.65 MeV with a width of 270 KeV.

The paper is organized as follows. In sect.~II we give a short outline of the LIT
calculation of the deuteron photodisintegration. In addition we discuss some aspects
of the asymptotic behavior  of the LIT solution $\tilde\Psi$. In sect.~III we describe the model 
for our fictitious NN interaction and show results for the corresponding $^3P_1$ 
phase shifts and the photoabsorption cross section calculated in the conventional
way with a $^3P_1$ scattering wave function. The LIT results of our case study are
discussed in sect.~IV.

\section{\label{LIT} The Formalism}

The deuteron total photoabsorption cross section is given by
\begin{equation}
\sigma_\gamma^d(\omega)= 4 \pi^2 \alpha \omega R^d(\omega)\,,
\end{equation}
where $\alpha$ is the fine structure constant, $\omega$ is the energy 
of the photon absorped by the deuteron, and $R^d(\omega)$ denotes
the response function defined as 
\begin{equation}\label{response}
   R^d(\omega)=\sum\!\!\!\!~\!\!\!\!\!\!\!\!\int _f\,\,
             |\langle f|\Theta| 0 \rangle |^2 \delta(\omega -E_{np}-E_d) \, .
\end{equation} 
Here $E_d$ and $|0\rangle$ are the deuteron bound state energy and wave function, $E_{np}$ 
and $|f\rangle$ denote relative kinetic energy and wave function of the outgoing
$np$ pair for a given two-nucleon Hamiltonian $H$, and $\Theta$ is the 
operator inducing the reaction (it is assumed that recoil effects are negligible). 
Here we consider the total photoabsorption cross section in unretarded dipole 
approximation, i.e.
\begin{equation}\label{operator}
\Theta=\sum_{i=1}^2 z_i \tau_i^3\,,
\end{equation}
where $z_i$ and  $\tau_i^3$ are the third components of position and isospin 
coordinates of the i-th nucleon.

For a conventional evaluation of $R^d(\omega)$ one determines the $np$ scattering 
wave functions for the induced NN partials waves ($^3P_0$, $^3P_1$, $^3P_2-^3F_2$)
and calculates the transitions matrix elements appearing in (\ref{response}).
It is evident that for each of the three partial waves a separate cross section
contribution can be defined.

As already mentioned, with the LIT method one avoids the explicit calculation of
scattering wave functions. Here one proceeds in the following way.
One first calculates the ground state wave function of the nucleus in question, in
our case the deuteron. Then one has to solve the equation
\begin{equation} \label{LITeq}
(H+E_d-\sigma_R - i \sigma_I)| \tilde \Psi \rangle = \Theta |0\rangle \,.
\end{equation}
Since $|\tilde\Psi\rangle$ is localized one needs to apply only a bound-state technique for the
solution of (\ref{LITeq}). It is convenient to perform multipole decompositions of left-
and right-hand sides of (\ref{LITeq}). In the here considered deuteron case this leads to three 
separate equations, each one for a different $\tilde\Psi_i$ (i=1,2,3). They correspond to the 
above mentioned three separate $R^d$ contributions due to the $^3P_0$, $^3P_1$, and $^3P_2-^3F_2$ 
$np$ final states. The calculation is carried out for many values of $\sigma_R$ and a fixed 
$\sigma_I$. A theorem based on the closure property of the eigenstates of $H$ shows that the
sum of the overlaps $\langle\tilde\Psi_i|\tilde\Psi_i\rangle$ of the three different
multipole solutions of (\ref{LITeq}) corresponds to the Lorentz integral transform of $R^d$,
\begin{equation}\label{deflit}
 {\rm L}^d (\sigma_R,\sigma_I)=\sum_{i=1}^3 \langle\tilde\Psi|\tilde\Psi\rangle= 
 \int R^d(\omega)\,  {\mathcal L}(\omega,\sigma_R,\sigma_I)\, d\omega\,,
\end{equation} 
where ${\mathcal L}$ is a Lorentzian centered at
$\sigma_R$ and with a width $\Gamma=2\sigma_I$:
\begin{equation}
{\mathcal L}(\omega,\sigma_R,\sigma_I)=
  {\frac {1} {(\omega -\sigma_R)^2 + \sigma_I^2}}\,.
\end{equation}
The parameters $\sigma_{R/I}$, for which (\ref{LITeq}) is solved, are chosen in relation
to the physical problem. In fact $\sigma_I$ represents a kind of energy resolution for the 
response function, while the values of $\sigma_R$ scan the region of interest.

In a final step the transform (\ref{deflit}) is inverted in order to obtain the response function
and thus the cross section. In the following we use the standard LIT inversion method  
(alternative inversion methods can be found in \cite{ALRS05}), however, with an extension that 
allows to take care of narrow resonances in the response function.

The standard LIT inversion method consists in the following ansatz for
the response function $R^d$:
\begin{equation}
R^d(E_{np}) = \sum_{m=1}^{M_{\rm max}} c_m \chi_m(E_{np},\alpha_i) \,,
\label{sumr}
\end{equation}
where we have replaced the argument $\omega$ of $R^d$ by $E_{np}=\omega-E_d$.
The $\chi_m$ are given functions with nonlinear parameters $\alpha_i$.
Here we take
\begin{equation}
\label{bset1}
\chi_1(E_{np},\alpha_i) = {\frac {1} {(E_{np} -E_{\rm res})^2 + ({\frac {\Gamma}{2}})^2}}
\left({\frac {1}{1+\exp(-1)}}-{\frac {1}{1+\exp((E_{np}-\alpha_3)/\alpha_3)}}\right)\,,
\end{equation}
\begin{equation}
\label{bsetn}
\chi_m(E_{np},\alpha_i) = E_{np}^{\alpha_4} \exp(- {\frac {\alpha_5 E_{np}} {m-1}}) 
\qquad {\rm for}\,\,\, m>1 \,,
\end{equation}
where in (\ref{bset1}) we have set $E_{\rm res}\equiv\alpha_1$ and $\Gamma\equiv\alpha_2$.
It is evident that $\chi_1$ represents a resonance of Lorentzian shape, where
the additional factor in brackets ensures that $R^d(E_{np})$ is zero for $E_{np}=0$.
For $m>1$ the set $\chi_m$ represents the usual basis function set for LIT inversions. 
Substituting such an expansion into the right hand side of
(\ref{deflit}) one obtains
\begin{equation}
{\rm L}(\sigma_R,\sigma_I) =
\sum_{m=1}^{M_{\rm  max}} c_m \tilde\chi_m(\sigma_R,\sigma_I,\alpha_i) \,,
\label{sumphi}
\end{equation}
where
\begin{equation}
\tilde\chi_m(\sigma_R,\sigma_I,\alpha_i) =
\int_0^\infty dE_{np} {\frac {\chi_m(E_{np},\alpha_i)} {(E_{np}-\sigma_R)^2 + \sigma_I^2}}
\,\,.
\end{equation}
For given values of $\alpha_i$ and $M_{\rm max}$ the linear parameters $c_m$ are determined from 
a best fit of L$^d(\sigma_R,\sigma_I)$ of (\ref{sumphi}) to the calculated
L$^d(\sigma_R,\sigma_I)$ of (\ref{deflit}) for a fixed $\sigma_I$ and a number of 
$\sigma_R$ points much larger than $M_{\rm max}$. In addition one should vary
the various nonlinear parameter $\alpha_i$ over a sufficiently large range. The parameter
$\alpha_4$, however, can in general be determined from the known threshold behavior of the
response function. In fact for our deuteron case one has transitions to $P$-waves
of the final state and thus $\alpha_4=3/2$. One starts the inversion by choosing 
a relatively low value of $M_{\rm max}$, e.g. $M_{\rm max}=6$, then one selects 
the overall best fit and repeats the procedure for increasing $M_{\rm max}$ 
up to the point that a stability of the inverted response is obtained and 
taken as inversion result. 

As pointed out above the key point of the LIT approach is the fact that one only works with
a function, $\tilde\Psi$, of finite norm. This localized function $\tilde\Psi$ contains
that information of the scattering process which is necessary to calculate reaction
cross sections. The asymptotic behavior of $\tilde\Psi$ is described by an exponential
fall-off with the argument $((2m/\hbar^2)\sigma_I)^{1/2}$, where $m$ is the nucleon mass and
$\sigma_I$ a parameter of the LIT method, which governs the resolution. The smaller $\sigma_I$
the better is the resolution and the more extended becomes $\tilde\Psi$.
In other words if one wants to resolve finer structures in the cross section
one has to ensure that $\sigma_I$ is small enough and that $\tilde\Psi$ is calculated
sufficiently far enough in the asymptotic region. 

In particular, in the following, we investigate up to which two-nucleon distance the localized 
function $\tilde\Psi$ has to be determined in order to obtain reliable results for the resonance
cross section. In addition we also try to simulate the conditions of a LIT calculation
for nuclei with more than two nucleons, where usually expansions on complete sets, e.g.
hypherspherical harmonic functions, are used. In these cases the solutions $\tilde\Psi$
are obtained in a kind of a spherical box of variable radius $R_{\rm max}$. Beyond
$R_{\rm max}$ $\tilde\Psi$ falls off rapidly. In order to simulate a similiar situation,
in our case study we use an asymptotic boundary condition for $\tilde\Psi$ leading to a
strong fall-off of $\tilde\Psi$ at a two-nucleon distance of $R_{\rm max}$ and study the
dependence of the results on $R_{\rm max}$.

\section{The potential model} 

In the following we will use the AV18 NN interaction \cite{AV18}, but modify the 
$^3P_1$ potential in order to introduce a fictitious resonance. This is achieved by
adding an attractive potential term, i.e. $V(^3P_1) \rightarrow V(^3P_1)+V_{\rm add}$.
We take
\begin{equation}
V_{\rm add} = -{\frac {57.6 {\rm MeV}}{r}}\left(1-\exp(-2r^2)\right)
\left(1+\exp({\frac{r-5}{0.2}})\right)^{-1}
\end{equation}
with the relative coordinate $r$ in units of fm.

In Fig.~1 we show the phase shift and in Fig.~2 the ``deuteron photoabsorption
cross section'' $\sigma_\gamma^d(^3P_1)$ for the modified $^3P_1$ potential
calculated with an explicit $^3P_1$ continuum wave function. The phase shift 
exhibits two resonances, one at $E_{np}$=0.48 MeV and a second one at about 10.5 MeV.
For $\sigma_\gamma^d(^3P_1)$ one finds a very pronounced and narrow resonance at  
$\omega$=2.65 MeV with a width $\Gamma$ of 270 KeV, while the second resonance 
is about four orders of magnitude weaker and has a width of about 5 MeV. There is
a third cross section peak for $\sigma_\gamma^d(^3P_1)$ at 60 MeV, about 25 times stronger
than the second resonance peak, with a rather large width, which cannot be ascribed
to a $^3P_1$ resonance.

\section{Results and discussion}

Before we investigate under which conditions one can reproduce the pronounced resonance 
of $\sigma_\gamma^d(^3P_1)$ of Fig.~2 via the LIT method, we first want to make another 
consideration. The smallest cross section width which has been resolved in previous 
LIT calculation is about 10 MeV (total photoabsorption cross sections of $^6$He \cite{6-body} 
and $^2$H \cite{jisp,ELOB07}). Thus, in order to better understand what was achieved in
such calculations, let us first consider the deuteron case with the unmodified AV18 potential. 
Since we are particularly interested how far the asymptotic range of $\tilde\Psi$ has to be 
taken into account we proceed for the solution of (\ref{LITeq}) as follows.
For a given value of $R_{\rm max}$ we use a boundary condition that requires a strong
fall-off of $\tilde\Psi$ at $R_{\rm max}$ and calculate the overlap 
$\langle\tilde\Psi|\tilde\Psi\rangle$ of (\ref{deflit}) only in the range  from $r=0$ to 
$r=R_{\rm max}$ (for more details of such a calculation see \cite{Stieltjes,ELOB07}). 
In Fig.~3 we show the convergence of L$^d(^3P_1)$ with respect to $R_{\rm max}$ taking, as 
for other previous LIT calculation of $\sigma_\gamma^d$ \cite{jisp,ELOB07}, $\sigma_I=10$ 
MeV. One sees that $R_{\rm max}$=10 fm is not sufficient, while $R_{\rm max}$=15 fm
leads already to a rather good approximation of the final result. Convergence is
essentially reached with $R_{\rm max}$=20 fm. One may ask whether $\sigma_I$=10 MeV
is sufficiently small to resolve the peak structure of $\sigma_\gamma^d$ in the inversion.
A minimal check can be performed in the following way. Let us assume that the
cross section is a $\delta$-peak at $E^{\rm peak}$ with size $\sigma^{\rm peak}$, i.e. 
\begin{equation}
\nonumber
\sigma_\gamma^d(E_{np})= \sigma^{\rm peak} \delta(E_{np}-E^{\rm peak}) \,.
\end{equation}
The resulting LIT is then given by the Lorentzian function 
\begin{equation}
\nonumber
{\rm L}^d_\delta(E^{\rm peak},\sigma_R,\sigma_I) = \sigma^{\rm peak} 
{\mathcal L}(E^{\rm peak},\sigma_R,\sigma_I) \,.
\end{equation}
In the right panel of Fig.~3 we show such an L$^d_\delta$, taking as $E^{\rm peak}$
the peak position of L$^d(R_{\rm max}=30$ fm) in the left panel of Fig.~3, in addition we set 
L$^d_\delta(E^{\rm peak})=$L$^d(E^{\rm peak})$. It is readily seen that L$^d$ has a 
considerably larger width than L$^d_\delta$ and thus should contain sufficient information 
of the peak structure for a reliable inversion. In fact in \cite{ELOB07} it was shown
that a LIT calculation of L$^d$ with $\sigma_I=10$ MeV leads precisely to the 
same cross section as a calculation with explicit $np$ continuum wave functions.

After these initial considerations we turn to $\sigma_\gamma^d(^3P_1)$ with the modified 
$^3P_1$ potential discussed in the previous section. We calculate L$^d(^3P_1)$ with various 
$\sigma_I$ values, namely $\sigma_I=10$, 5, 2, 1, 0.5, and 0.1 MeV. First we discuss the 
convergence of the LIT with respect to $R_{\rm max}$ at low energies. In Figs.~4 and 5
we show the cases with $\sigma_I=1$ and 0.5 MeV as examples. Different from the 
case considered in Fig.~3, $R_{\rm max}$=20 fm is not sufficient to reach a good convergence. 
For $\sigma_I=1$ MeV one finds that the result for $R_{\rm max}$=30 fm differs up to about 
1\% from the converged LIT showing in addition a somewhat oscillatory behavior. With 
$R_{\rm max}$=40 fm one sees a much smoother curve and deviations are reduced to about 0.5\%. 
Very good convergence is reached with $R_{\rm max}$=50 fm with deviations of about 0.1\%.
A rather similar trend can be observed in Fig.~5 for $\sigma_I=0.5$ MeV, but the deviations 
from the converged result are larger, e.g. for $R_{\rm max}$=30 fm one has deviations up 
to 8\%, and the oscillatory behavior becomes more pronounced and is even visible for 
$R_{\rm max}$=50 fm. For larger $\sigma_I$ values (2, 5, 10 MeV) one finds a similar
situation as shown in Figs.~4 and 5, but  with increasing $\sigma_I$ relative deviations 
become smaller and oscillations tend to vanish. For $\sigma_I=0.1$ MeV it is not sufficient 
to take $R_{\rm max}$=80 fm in order to have a converged result. This is illustrated in 
Fig.~6. One observes a steady decrease of the peak height up to an $R_{\rm max}$ of almost 
140 fm. In order to obtain a smooth result in the whole peak region one has to further
increase $R_{\rm max}$ up to 300 fm. It is interesting to observe in Fig.~6 how this 
convergence is reached with growing $R_{\rm max}$. For values of 30, 60 and 90 fm two, 
three, and four separate Lorentzians, respectively, are visible. For even greater $R_{\rm max}$
the density of single Lorentzians increases further leading finally to a smooth curve
for $R_{\rm max}$=300 fm. The general picture looks very similar to LIT results obtained by
solving $\tilde\Psi$ via an expansion on a complete basis (see e.g. \cite{ELOB07}),
there an increase of basis states has a similarly effect on existing oscillations
as a growing $R_{\rm max}$ in our case.

As next point we want to check which value of $\sigma_I$ should be sufficient to obtain good 
inversion results at low energies, i.e. leads to a good description of the pronounced resonance 
peak in the cross section. For this purpose we perform a similar study as has been illustrated 
in the right panel of Fig.~3. In Fig.~7 we show the corresponding results for L$^d$ and 
L$^d_\delta$ for the various considered $\sigma_I$ values. For $\sigma_I=10$ MeV both 
results are essentially identical and for $\sigma_I=5$ MeV they are almost identical. A 
further reduction of $\sigma_I$ leads to more and more differences between L$^d$ and 
L$^d_\delta$. Finally, for $\sigma_I=0.1$ MeV L$^d$ has a much larger width than L$^d_\delta$. 
Thus one can be rather sure that a very good inversion result should be obtained for the
pronounced low-energy resonance with our lowest $\sigma_I$ value, while $\sigma_I$=10 MeV 
should not be sufficient for a reliable inversion. For the other $\sigma_I$ values it is 
not clear beforehand whether they suffice, and in the following this will be investigated 
in detail.

At first, in Fig.~8 we show the results of a minimal inversion, where in (\ref{sumr}) we use 
$M_{\rm max}$=1 and hence allow only for a single resonance in the cross section. It is 
evident that with decreasing $\sigma_I$ the resonance peak is steadily reduced and the 
resonance width steadily increased. In the right panel of Fig.~8 a comparison is made between 
the $\sigma_I$=0.1 MeV result and the cross section calculated in the conventional way with 
a $^3P_1$ $np$ continuum wave function. One finds an excellent agreement in the peak region.
Thus it seems that it is necessary to use (i) large $R_{\rm max}$ values of far more than 100 fm
and (ii) $\sigma_I$ values equal or smaller than the resonance width $\Gamma$. However,
in the further discussion we will illustrate that, fortunately, this is actually not the case.

Before we come to the discussion of more complete inversion results, in Fig.~9 we first show 
the LIT L$^d$ at higher energies considering three $\sigma_I$ and various $R_{\rm max}$ values. 
For $\sigma$=5 MeV and $R_{\rm max}$=30 fm one finds slight oscillations beyond 40 MeV, while 
a larger $R_{\rm max}$ leads to a smooth curve. The picture changes for $\sigma$=1 MeV, 
oscillations are present in all the three illustrated cases and start already beyond 20 MeV, 
it is also seen that the oscillation frequency (amplitude) increases (decreases) with growing 
$R_{\rm max}$. For $\sigma$=0.1 MeV we only show the result with $R_{\rm max}$=300 fm. One 
notes that strong oscillations are present beyond 20 MeV.

As next point we turn to the full inversion of L$^d$, i.e. not using as for the results of
Fig.~8 a restriction of $M_{\rm max}$ to 1, but allowing for higher $M_{\rm max}$ values. We 
first consider $\sigma_I$=1 MeV and take various $R_{\rm max}$ values. With $R_{\rm max}$=20 fm 
we do not obtain reasonable results, since the pronounced low-energy resonance cannot be 
reproduced, instead the best inversion fits are found for a vanishing width $\Gamma$ of the 
resonance, which means that a $\delta$-peak is preferred. As illustrated in Fig.~10 the 
situation changes for $R_{\rm max}$=30 fm. For the low-energy region one has a very stable 
inversion result that starts already with $M_{\rm max}$=7 and leads to a very good description 
of the resonance peak. At higher energies the picture is different. None of the inversions 
describes the region of the second resonance correctly. At even higher energy inversions are 
improving with increasing $M_{\rm max}$ but without reaching a very good agreement with the 
conventional calculation. The reason for not reproducing the high-energy cross section more 
precisely is due to the fact that the low-energy resonance contains almost all the E1($^3P_1$)
strength. Even for a high-energy $\sigma_R$ the by far dominant LIT contribution stems from the
low-energy resonance region, so that a precise inversion of the high-energy region requires also
a very accurate calculation of the LIT. Similar results as for $\sigma_I=1$ MeV are also found
for $\sigma_I=2$ MeV, while our other $\sigma_I$ values do not lead to a sufficiently good 
description of the low-energy resonance if $R_{\rm max}$ is taken equal to 30 fm. For  
$R_{\rm max}$=50 fm, also shown in Fig.~10, it does not come as a surprise that the low-energy 
region is desribed again in an excellent way. In addition for the higher $M_{\rm max}$ values 
one finds a rather good description of the second resonance region and also the high-energy 
cross section is sufficiently accurate. In Fig.~11 we show results with $R_{\rm max}$=80 fm 
and $\sigma_I$=0.5, 2, and 5 MeV. The smaller two $\sigma_I$ values lead to excellent 
descriptions of $\sigma_\gamma^d(^3P_1)$ in the whole considered energy range, only that with 
$\sigma_I$=0.5 MeV the cross section is underestimated beyond 120 MeV, while the $\sigma_I$=5 MeV 
result exhibits a to 5\% too low resonance peak cross section. We do not show inversion 
results with $\sigma_I$=0.1 and 10 MeV. They are rather unstable and show oscillations with 
rather pronounced unphysical negative cross sections.

We summarize our results as follows. We have performed a case study for a LIT calculation 
with a pronounced resonance in the cross section. To this end the nucleon-nucleon interaction 
has been modified in the $^3P_1$ partial wave to generate a fictitious low-lying 
resonance in the $np$ continuum. The resulting ``deuteron photoabsorption cross section''
has been calculated in two ways: (i) with an explicit $^3P_1$ $np$ continuum wave function
and (ii) via the LIT method, i.e. calculating the Lorentz integral transform with subsequent
inversion in order to obtain the cross section. The conventional calculation leads to a
cross section with three structures: (i) a very pronounced resonance with width of 270 KeV 
and a peak position of 0.48 MeV above the deuteron break-up threshold, (ii) a second resonance 
at 10 MeV above threshold with a width of about 5 MeV, which is four orders of magnitude weaker 
than the first one,
and (iii) a rather broad maximum at a photon energy of about 60 MeV with a 25 times 
higher peak than the second resonance maximum. For a proper resolution of the dominant 
cross section structure, the pronounced low-energy resonance, one has to consider rather
long-ranged solutions $\tilde\Psi$ of the LIT equation, namely at least up to $R_{\rm max}=30$ 
fm. The cross section results depend strongly also on the LIT parameter $\sigma_I$, which 
governs the resolution (the smaller $\sigma_I$ the better the resolution). It turns out that 
it is advantageous to work with $\sigma_I=1$ and 2 MeV, about 4-8 times larger than $\Gamma$,
in this case $R_{\rm max}=$30 fm is sufficient. Even larger $\sigma_I$ do not correctly
reproduce the resonance cross section ($\sigma_I$=5 MeV) or even lead to a $\delta$-shape 
resonance ($\sigma_I$=10 MeV). Very small $\sigma_I$ values also lead to disadvantages, 
since a larger $R_{\rm max}$ is required, e.g. with $\sigma_I$=0.1 MeV, even taking
$R_{\rm max}=300$ fm the LIT is only poorly converged for energies beyond the first resonance 
region making inversions completely unreliable. Using the intermediate $\sigma_I$ values 
and setting $R_{\rm max}=30$ fm one obtains besides the excellent reproduction of the
low-energy resonance also a rather good description of the broad maximum at higher energies, 
while the second, extremely weak, resonance is not well described. An increase
of $R_{\rm max}$ to 50 fm leads to a rather good description also of the tiny second resonance
cross section, and with $R_{\rm max}=80$ fm there is further improvement so that an excellent
agreement with the conventional calculation is achieved in the whole considered energy range
if $\sigma_I$ is chosen between 0.5 to 2 MeV.

One might think that for a good description of the low-energy resonance it could
be sufficient to restrict the LIT inversion such that only a single resonance
is allowed in the cross section. We have shown that this is actually not the case. Even taking
$R_{\rm max}=80$ fm one obtains results which show an overestimation of the peak height
and an underestimation of the width, e.g. with $\sigma_I=5$ MeV (1 MeV) one
obtains a width of 90 KeV (180 KeV) instead of 270 KeV and a peak height which is
overestimated by a factor of 2.7 (1.4). In addition the resonance is slightly shifted to 
higher energies. Only for $\sigma_I=0.1$ MeV and $R_{\rm max}=300$ fm one finds a very
good result for the resonance region. Fortunately, as pointed out in the previous paragraph,
the picture changes if one performs a full LIT inversion, i.e. allows for structure besides
the resonance.  

We think that these findings about the LIT resolution and the proper consideration of the
asymptotic extension of the LIT solution $\tilde\Psi$ are generally important for LIT 
calculations. In particular for methods where $\tilde\Psi$ is expanded on a complete set 
it has to be ensured that the solution extends far enough into the asymptotic region
if one wants to properly resolve cross section structures with a small width.
If this is guaranteed the LIT approach opens the possibility to calculate also narrow cross
section resonances with great precision.

\newpage


\begin{figure}[ht]
\resizebox*{16cm}{7cm}{\includegraphics*[angle=0]{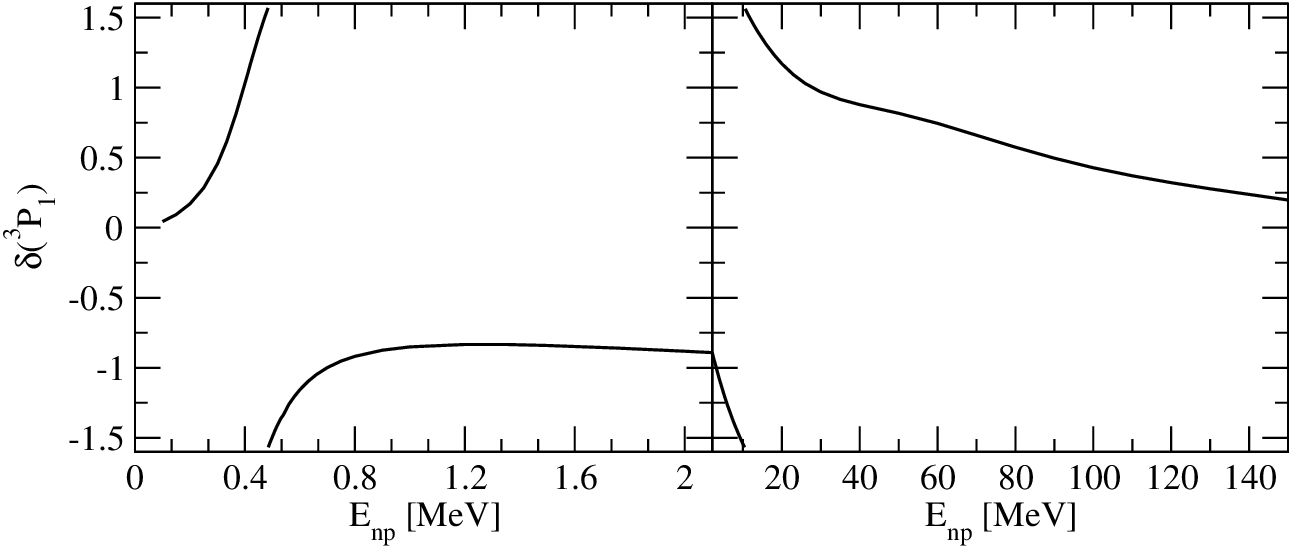}}
\caption{Phase shift $^3P_1$ with modified $V(^3P_1)$ potential at low energy (left) and 
for extended energy range (right).
}
\end{figure}

\begin{figure}[ht]
\resizebox*{16cm}{7cm}{\includegraphics*[angle=0]{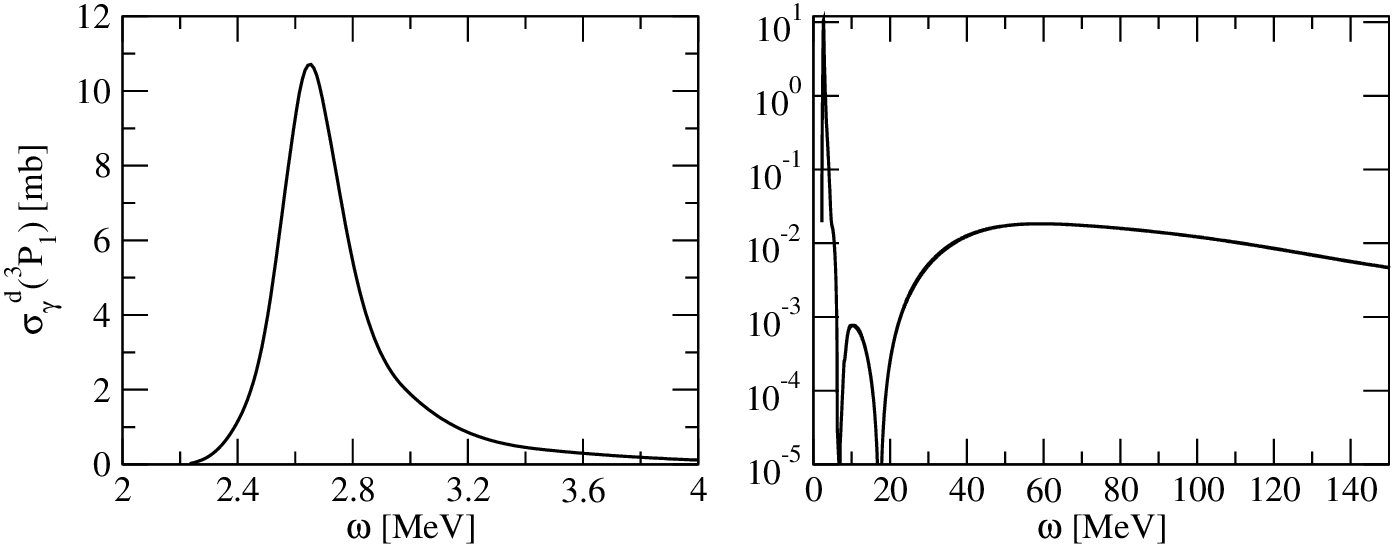}}
\caption{Photodisintegration cross section $\sigma^d_{\gamma}(^3P_1)$ with modified 
$V(^3P_1)$ potential at low energy (left) and for extended energy range (right).
}
\end{figure}

\begin{figure}[ht]
\resizebox*{16cm}{7cm}{\includegraphics*[angle=0]{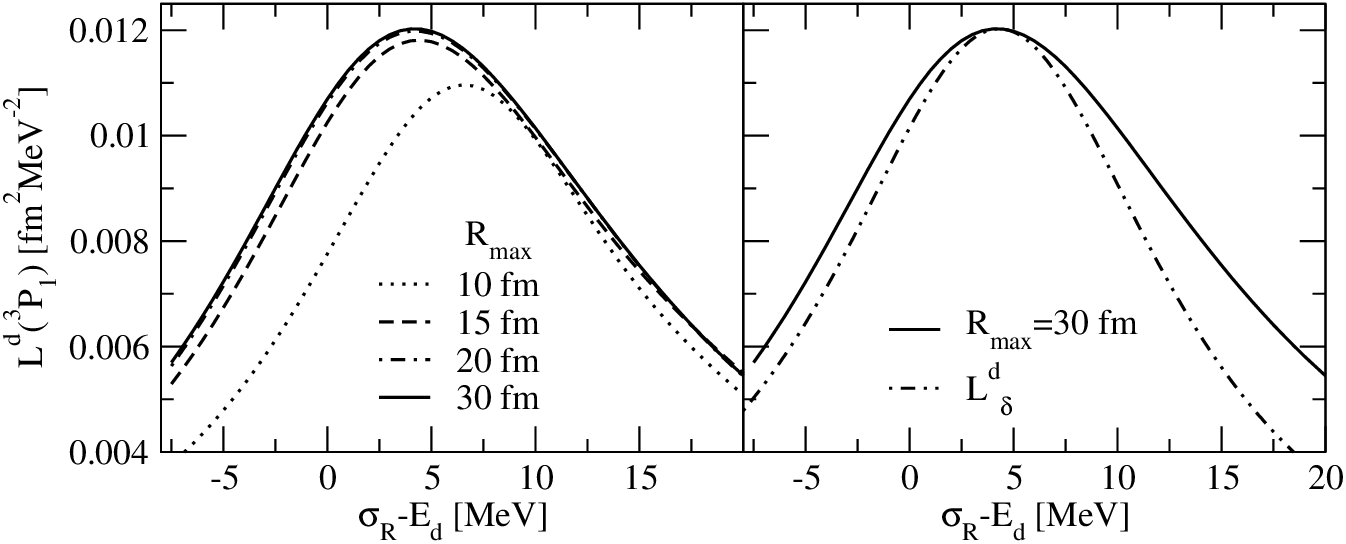}}
\caption{Lorentz integral transform L$^{\rm d}(^3P_1)$ with AV18 NN potential in low-energy
region for $\sigma_I=10$ MeV and various values of $R_{\rm max}$ (left) and comparison of 
the $R_{\rm max}=30$ fm result with L$^{\rm d}_\delta$ (right).
}
\end{figure}

\begin{figure}[ht]
\resizebox*{6.5cm}{10.5cm}{\includegraphics*[angle=0]{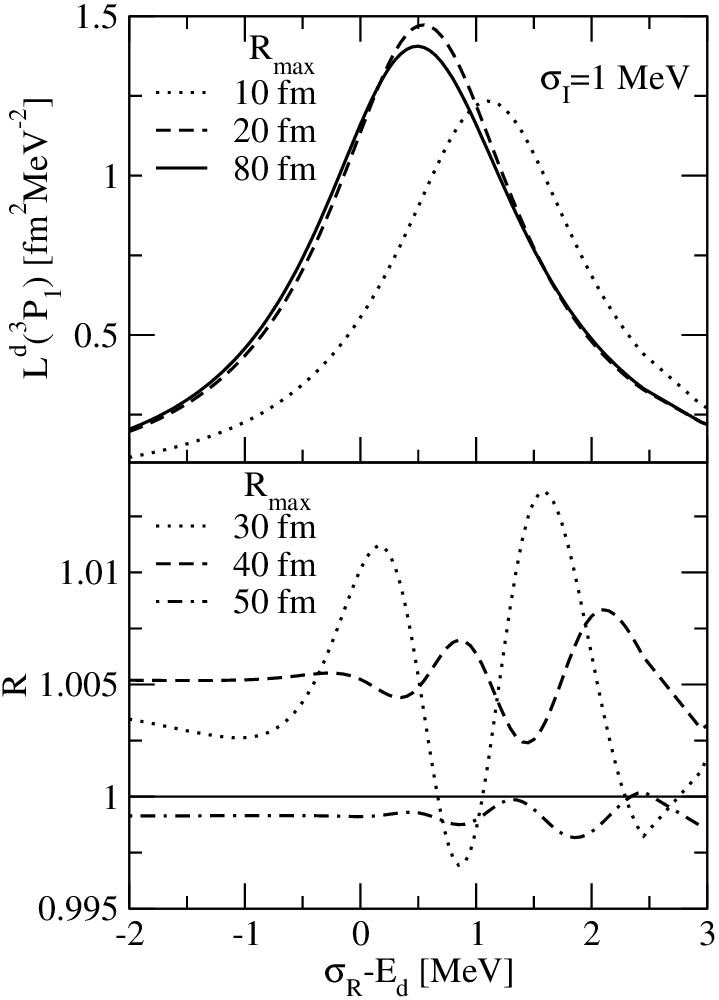}}
\caption{Lorentz integral transform L$^{\rm d}(^3P_1)$ with modified $V(^3P_1)$ potential in
first resonance region for $\sigma_I=1$ MeV and various values of $R_{\rm max}$ (top) and ratio 
R=L$^{\rm d}(R_{\rm max})/L^{\rm d}(R_{\rm max}=80$ fm) (bottom).
}
\end{figure}

\begin{figure}[ht]
\resizebox*{6.5cm}{10.5cm}{\includegraphics*[angle=0]{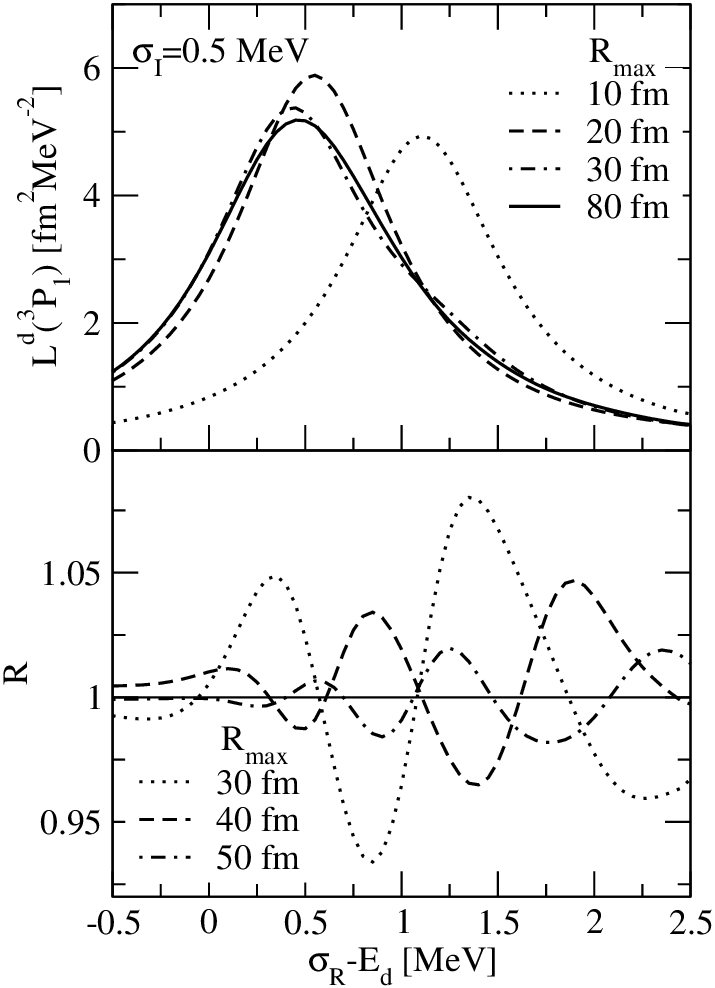}}
\caption{As Fig.~4 but for $\sigma_I=0.5$ MeV 
}
\end{figure}

\begin{figure}[ht]
\resizebox*{16cm}{7cm}{\includegraphics*[angle=0]{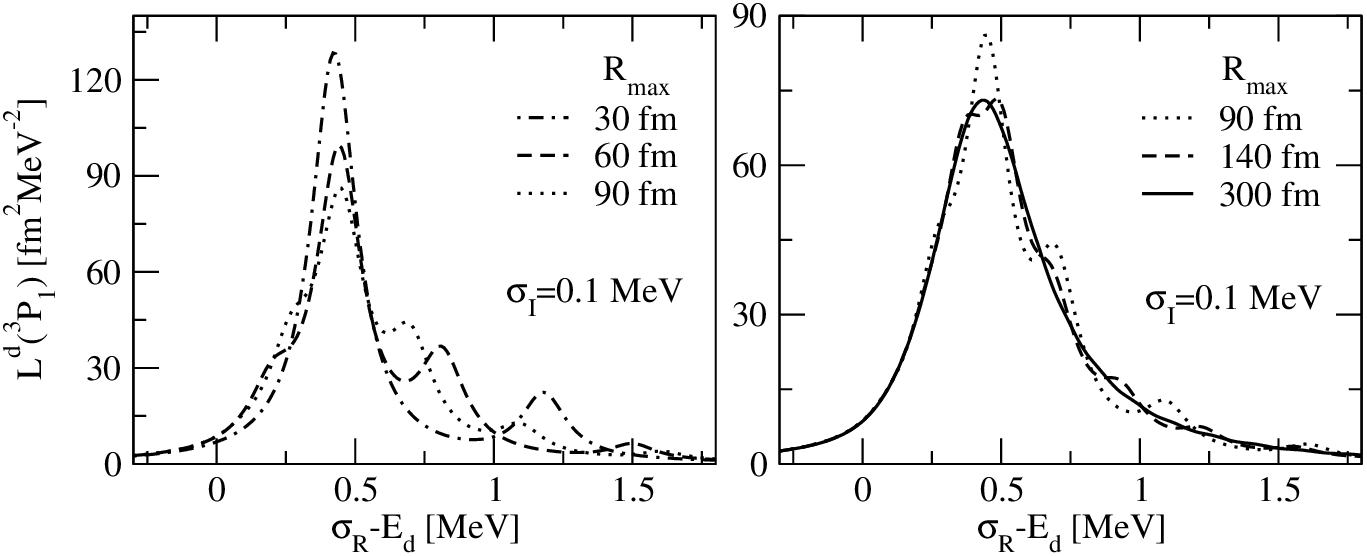}}
\caption{As Fig.~4 but for $\sigma_I=0.1$ MeV and various values of $R_{\rm max}$ 
in both panels.
}
\end{figure}

\begin{figure}[ht]
\resizebox*{16cm}{13cm}{\includegraphics*[angle=0]{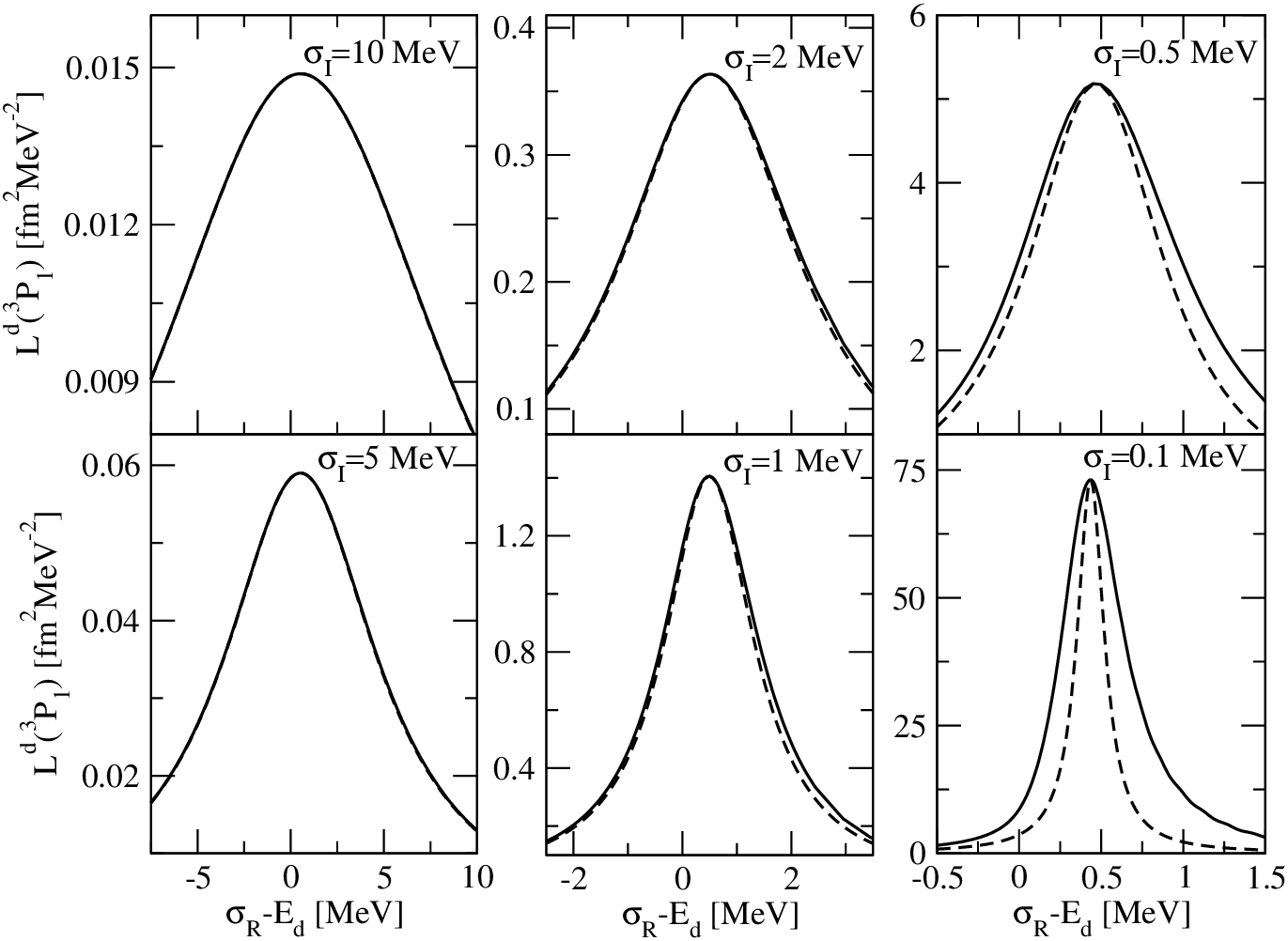}}
\caption{Lorentz integral transform L$^{\rm d}(^3P_1)$ (full curves) with modified $V(^3P_1)$ 
potential in first resonance region in comparison to L$^{\rm d}_\delta$ (dashed curves) for 
various values of $\sigma_I$ 
with $R_{\rm max}=80$ fm, except for $\sigma_I=0.1$ MeV where $R_{\rm max}$ is equal to 300 fm.  
}
\end{figure}

\begin{figure}[ht]
\resizebox*{16cm}{7cm}{\includegraphics*[angle=0]{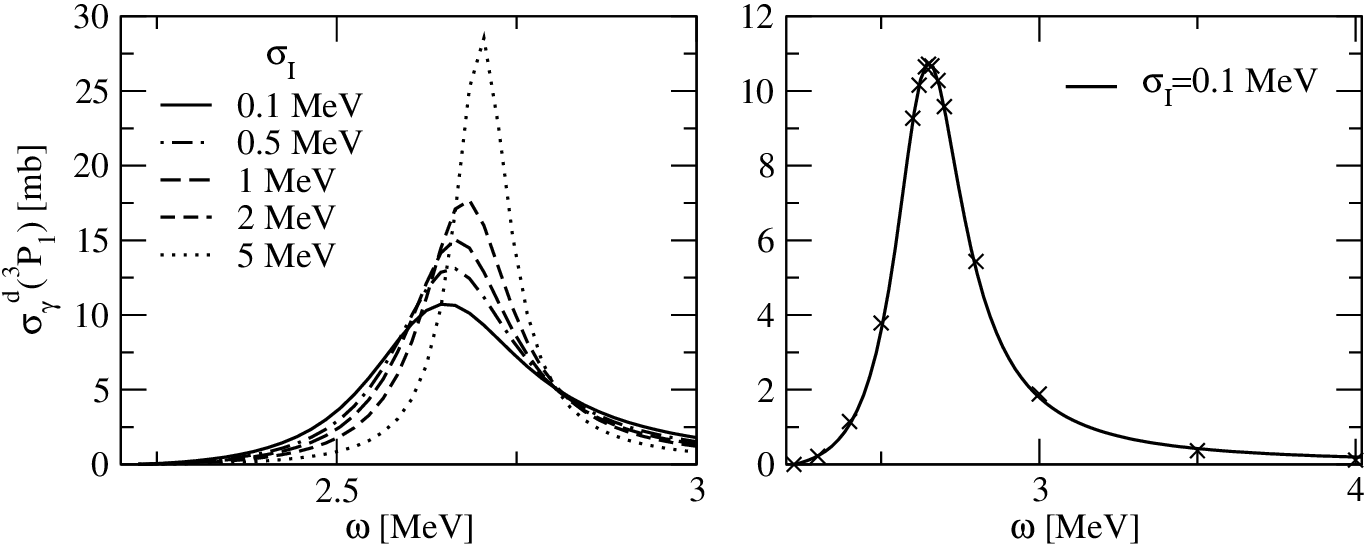}}
\caption{Cross section $\sigma^d_{\gamma}(^3P_1)$ with modified $V(^3P_1)$ potential in first
resonance region obtained from ``minimal'' inversion with $M_{\rm max}=1$ (see text) for 
various $\sigma_I$ values (left) and comparison of the $\sigma_I=0.1$ MeV result (full curve) 
with result of conventional calculation with explicit $np$ continuum wave functions (crosses). 
}
\end{figure}

\begin{figure}[ht]
\resizebox*{7cm}{10cm}{\includegraphics*[angle=0]{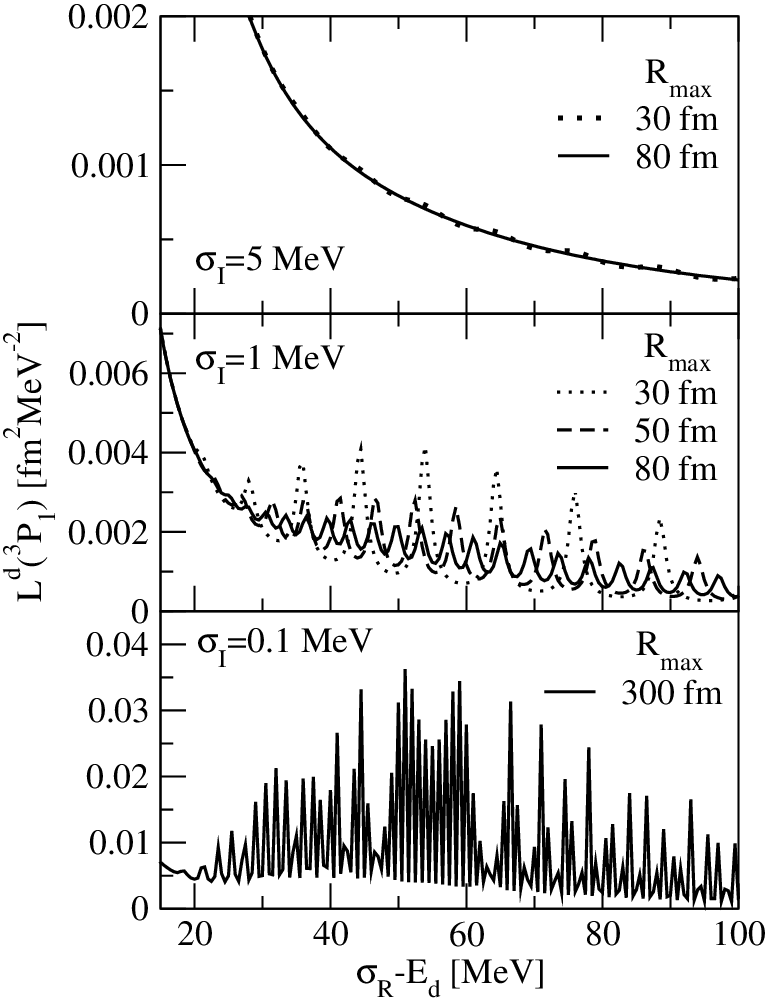}}
\caption{Lorentz integral transform L$^{\rm d}(^3P_1)$ with modified $V(^3P_1)$ potential
beyond first resonance region for $\sigma_I=5$ MeV (top), 1 MeV (middle), and 0.1 MeV 
(bottom) and various values of $R_{\rm max}$.
}
\end{figure}

\begin{figure}[ht]
\resizebox*{13cm}{9.5cm}{\includegraphics*[angle=0]{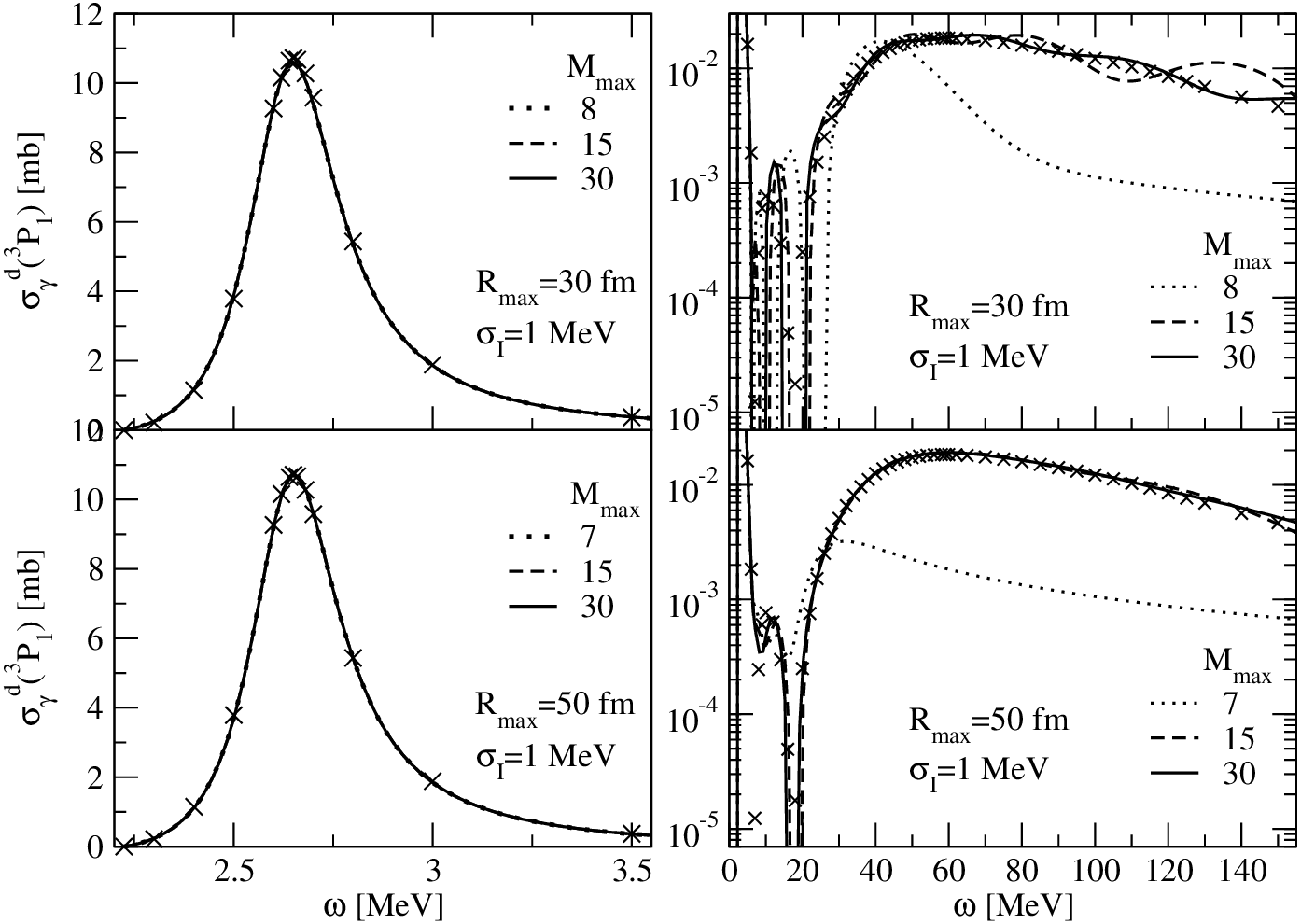}}
\caption{Cross section $\sigma^d_{\gamma}(^3P_1)$ with modified $V(^3P_1)$ potential
obtained from inversion of L$^{\rm d}(^3P_1, \sigma_I=1$ MeV, $R_{\rm max})$ with various values 
of $M_{\rm max}$ and for $R_{\rm max}=30$ fm (top) and 50 fm (bottom) in first resonance 
region (left) and at higher energies (right); also shown
results of conventional calculation with explicit $np$ continuum wave functions (crosses).
}
\end{figure}

\begin{figure}[ht]
\resizebox*{16cm}{7cm}{\includegraphics*[angle=0]{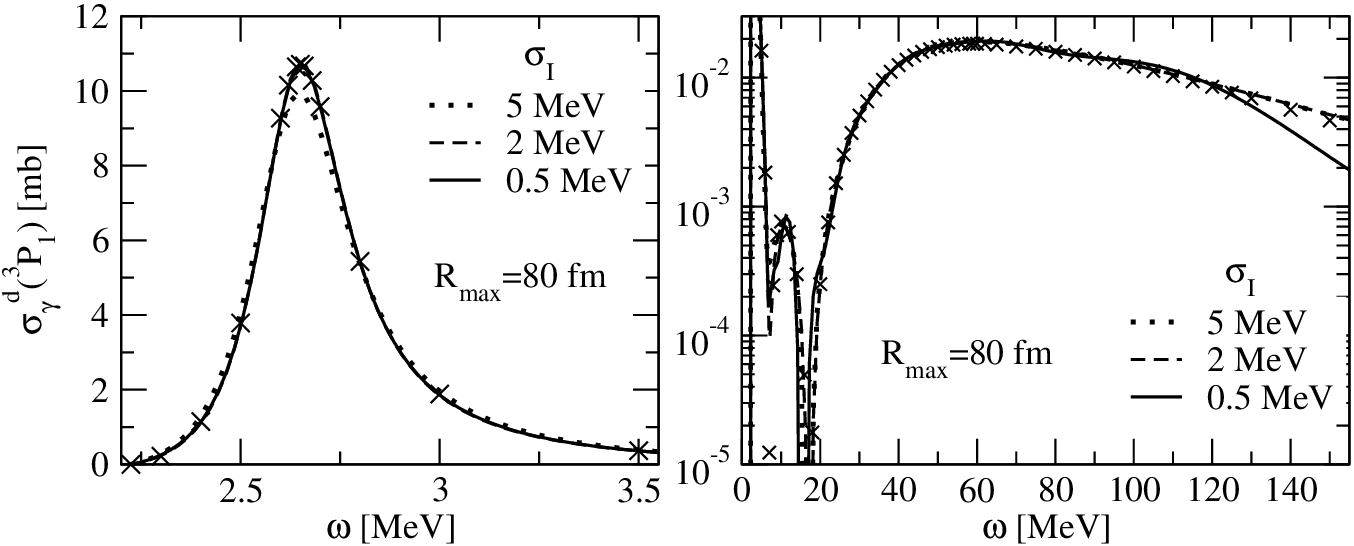}}
\caption{As Fig.~10 but for $R_{\rm max}=80$ fm and $\sigma_I=5$, 2, and 0.5 MeV.
}
\end{figure}

\end{document}